\documentclass[seceq]{ptptex}

\usepackage{graphicx}
\usepackage{wrapft}

\title{%        %You can use \\ for explicit line-break
Is our Universe brany ?%
}

\author{%       %Use \scshape  for the family name
David \textsc{Langlois}%
}

\inst{%     %Affiliation, neglected when [addenda] or [errata]
APC\ (Astroparticules et Cosmologie)
\footnote{UMR 7164 (CNRS, Universit\'e Paris 7, CEA, Observatoire de Paris)}
,\\
 11 Place Marcelin Berthelot, 75005 Paris, France\\
 and\\
GRECO, Institut d'Astrophysique de Paris, \\ 
98bis Boulevard Arago, 75014 Paris, France\\
}

\abst{%         %this abstract is neglected when [addenda] or [errata]
In brane-worlds, our universe is assumed to be 
a submanifold, or brane, embedded in a higher-dimensional 
bulk spacetime. Focusing on scenarios with a curved five-dimensional 
bulk spacetime, I discuss their gravitational and cosmological properties. 
}

\begin{document}

\def\E{E}
\def\H{E}
\def\d{\delta}
\def\A{A}
\def\M{M_5}
\def\bi{\begin{itemize}}
\def\ei{\end{itemize}}
\newcommand{\beq}{\begin{equation}} 
\newcommand{\eeq}{\end{equation}}
\newcommand{\bea}{\begin{eqnarray*}} 
\newcommand{\eea}{\end{eqnarray*}}
\def\C{{\cal C}}
\def\T{T^{\rm (bulk)}}
\newcommand{\gsim}{\ \raise.3ex\hbox{$>$\kern-.75em\lower1ex\hbox{$\sim$}} \ } 
\newcommand{\lsim}{\ \raise.3ex\hbox{$<$\kern-.75em\lower1ex\hbox{$\sim$}} \ }

\maketitle

\section{Introduction}

Brane-world models, which have received a lot of
 attention during the last few years, are 
essentially characterized by two basic ideas:
\bi
\item they assume the existence of spatial extra dimensions, in addition to our four 
space-time dimensions. The higher dimensional spacetime is usually called the ``bulk''
spacetime;
\item  our accessible Universe is assumed to be a submanifold, called ``brane'', embedded in 
the bulk spacetime. Ordinary matter fields are assumed to be confined on the brane.
\ei
This confinement of matter on a submanifold is really the  novel ingredient that distinguishes brane-worlds from the more ancient models with extra-dimensions, based on the ideas 
of Kaluza and Klein. 

The motivations for studying brane-worlds are multiple. First, branes appear in string/M theory.
One example is the D-branes, corresponding to 
solitonic objects where open strings end.  Another example is  the two 
end-of-the-world branes of the Horawa-Witten model. 
Another  interest is that some of the brane-world models
have a strong link with the AdS/CFT correspondence.
Brane-worlds have also turned to be very fruitful to address various questions of 
particle physics, in particular the hierarchy problem. 
Finally, brane-worlds are also particularly interesting for their gravitational properties.  
 
In part because of this multiplicity of motivations, there exist many models of brane-worlds. 
It is often convenient to regroup them in two broad categories: 
\bi 
\item  models with compact flat extra dimensions
\item models with warped extra-dimensions.
\ei
Another possible 
classification would be to distinguish models with so-called TeV gravity, i.e. for which 
the fundamental Planck mass is of the order of the TeV, from other models with a higher 
fundamental Planck mass.

 In this contribution, I will  focus my attention on 
 brane-worlds characterized by 
 a single extra  dimension   where  the bulk space-time is 
{\it curved} instead of flat and where 
the  self-gravity of the brane is  taken into account. 
This includes the 
configurations discussed by Randall and Sundrum. The present contribution is limited to 
a few selected topics. Other aspects and more details can  
can be found in several introductory reviews \cite{reviews}.   

\section{Gravity in the brane}
The main constraint on brane-worlds models is to recover usual gravity, at least approximately,
in our brane-universe. Whereas the usual trick is to compactify the extra dimensions on 
a sufficiently small scale, another possibility was emphasized  by Randall and Sundrum\cite{rs99b}, 
who considered  {\it curved}, or {\it warped}, 
bulk geometries. They realized
that  {\it compact} extra-dimensions are not necessary  to obtain a 
four-dimensional behaviour because 
the bulk curvature can  lead to an 
{\it effective compactification}.

\subsection{The Randall-Sundrum model}
The (second) Randall-Sundrum\cite{rs99b}
model is based on the following ingredients:
\bi
\item  a five-dimensional
bulk spacetime,  empty, but endowed with a negative cosmological constant
\beq
\Lambda=-{6\over\ell^2},
\eeq
\item a self-gravitating brane, which represents our world, 
endowed with a tension $\sigma$, 
and assumed to be $	Z_2$-symmetric. 
\ei
The five-dimensional Einstein equations are given by 
\beq
G_{AB}+\Lambda g_{AB}=\kappa^2 T_{AB}, 
\eeq
where $\kappa^2$ is the gravitational  coupling. The corresponding
five-dimensional Planck mass, $M_5$, can be defined as
\beq
  \kappa^2=M_5^{-3}.
\eeq
The bulk being empty, only the brane contributes to the energy-momentum
tensor $T_{AB}$. 
There are two equivalent ways of  solving Einstein's equations. Either
one solves them  directly by taking into account the presence of the
brane, assumed to be infinitely thin along the extra-dimension, 
in the form of a {\it distributional} energy-momentum tensor. Or, one 
solves first 
 the {\it vacuum} Einstein equations, i.e. setting the right hand side 
to zero, and then, one  takes into account the brane by imposing  
appropriate {\it junction conditions} at the spacetime boundary 
where the brane is located. These boundary  conditions are the 
generalization, to five dimensions, of the so-called Israel (-Darmois)
junction conditions and read 
\beq
\label{junction}
[K_{AB}]=-\kappa^2\left(T_{AB}-{T\over 3}h_{AB}\right).
\eeq
They relate the jump, between the two sides of the brane, of 
 the  extrinsic curvature tensor, defined by 
$K_{AB}\equiv h_A^C{D_C n_B}$ (where $n^A$ is the unit vector normal to the
brane and $h_{AB}=g_{AB}-n_An_B$ is the induced metric on the brane), to 
the brane energy-momentum 
tensor. For a $Z_2$ symmetric brane, the jump of the extrinsic curvature
is simply twice the value of the extrinsic curvature on one side of the 
brane.

Provided the tension satisfies the {\it constraint} 
\beq
\label{rs_constraint}
\Lambda+{\kappa^4\over 6}\sigma ^2=0,
\eeq
which implies in particular that $\sigma= 6\M^3/\ell$, it can be 
shown that the five-dimensional Einstein equations admit  the 
following {\it static} solution
\beq
\label{rs}
ds^2=a^2(y)\eta_{\mu\nu} dx^\mu dx^\nu+dy^2,
\eeq
\begin{figure}[t]
  \begin{center}
    \includegraphics[height=10pc,angle=0]{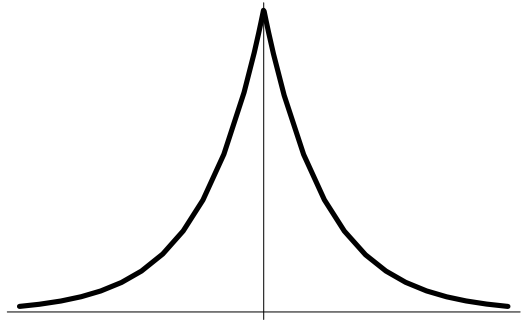}%%%%% ``13pc'' is just the example.
  \end{center}
\caption{Warping factor $a(y)$.}
\end{figure}
where $\eta_{\mu\nu}$ is the usual Minkowski metric and 
$a(y)$ is a {\it warping} scale factor, whose explicit dependence 
on $y$ is given by
\beq
\label{a}
a(y)=e^{-|y|/\ell},
\eeq
as shown on Fig.~1. 
Here, the brane is located at $y=0$ and the $Z_2$ symmetry means
that  $y$ with $-y$ are identified.
This bulk solution (\ref{rs}-\ref{a}) 
can also be interpreted as two identical portions 
of AdS (Anti-de Sitter) spacetime glued together at the brane location.
 
\subsection{Gravity in the Randall-Sundrum model}
Let us now investigate the effective gravity in this 
model, as measured by  an observer located 
on the brane. 
A first, and rather simple, step is to compute the effective four-dimensional
Planck mass. This can be done by substituting in the five-dimensional 
Einstein-Hilbert action,
\beq
 S_{\rm grav}={\M^3\over 2}\int d^4x \, dy \, \sqrt{-g}\, R, 
\eeq
the metric (\ref{rs}) and by integrating over the extra-dimension. One then identifies the 
factor in front of the resulting four-dimensional 
Einstein-Hilbert action (for $\eta_{\mu\nu}$)  with $M_{Pl}^2/2$, which gives 
\beq
\label{planck}
M_{Pl}^2=\M^3 \int_{-\infty}^{+\infty} dy\  a^2(y)= \M^3\ell.
\eeq
It is important to emphasize that the extra-dimension extends here to 
infinity. In the absence of the warping factor $a(y)$ this would lead 
to an infinite four-dimensional Planck mass. The warping of the 
extra-dimension, governed by the AdS lengthscale $\ell$, thus leads to 
an {\it effective compactification}, even if the extra-dimension is 
infinite.

To explore further the gravitational behaviour and derive 
for example the effective potential of a point mass located on the brane, 
one must study the  perturbations about the background metric 
(\ref{rs}).
Perturbing the metric, $g_{AB}=\bar g_{AB}+{h_{AB}}$, 
and working in the gauge $h_{yy}=0$, $h_{y\mu}=0$, $h_\mu^\mu=0$, 
$\partial_\mu h^\mu_\nu=0$, one finds that the linearized Einstein 
equations reduce to 
\beq
\left(a^{-2}\partial^2_{(4)}+\partial_y^2-{4\over \ell^2}+{4\over \ell}
\delta(y)\right)h_{\mu\nu}=0.
\eeq
This equation is separable and the solutions can be written as 
the superposition of eigenmodes $h(x^\mu,y)=u_m(y)
e^{ip_\mu x^\mu}$, with $p_\mu p^\mu=-m^2$. The 
dependence on the fifth dimension of the  modes is governed 
by  a {\it Schr\"odinger-like equation}:
\begin{figure}[t]
  \begin{center}
    \includegraphics[height=12pc,angle=0]{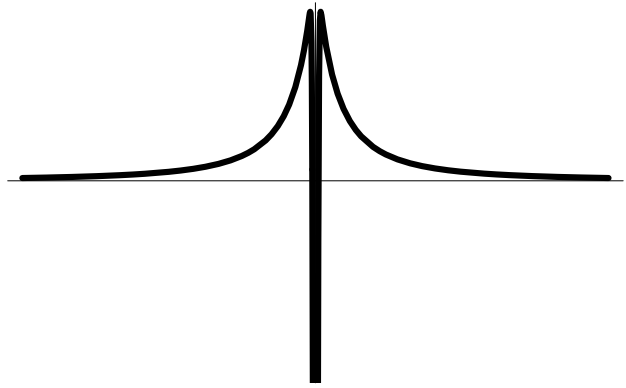}
  \end{center}
\caption{Effective potential of the Schr\"odinger-like equation that governs the dependence on the fifth dimension.}
\end{figure}
\beq
{d^2{\psi}_m\over dz^2}-V(z){\psi}_m=-m^2{\psi}_m,
\qquad
V(z)={15\over 4(|z|+\ell)^2}-{3\over \ell}\delta(z),
\eeq
where $\psi_m=a^{-1/2}u_m$ and  $z=\int dy/a(y)$.
The potential $V(z)$, plotted in Fig.~2, 
 is ``volcano''-shaped and 
goes to zero at infinity.

One can divide the 
solutions of this Schr\"odinger-like equation into: 
\bi
\item a zero mode ($m=0$),
$u_0(y)=a^2(y)/\sqrt{\ell}$,
which is concentrated near the brane and reproduces  the usual 
behaviour of 4D gravity;
\item a continuum of massive modes ($m>0$), which are weakly coupled to the 
brane and  modify standard 4D gravity.
\ei
More specifically, the perturbed metric 
outside a spherical source of mass $M$, and for $r\gg \ell$, is given by
\cite{gt} 
\beq
{\bar h}_{00}\simeq {2GM\over r}\left(1+{2\ell^2\over 3r^2}\right),\quad 
{\bar h}_{ij}\simeq {2GM\over r}\left(1+{\ell^2\over 3r^2}\right),
\eeq
where the bar here means that the perturbations are expressed  in 
the Gaussian
Normal gauge (i.e. $h_{yy}=h_{y\mu}=0$ and the brane is located at $y=0$) and 
thus correspond directly to the quantities measured on the brane. 
Standard gravity is thus recovered on scales $r\gg \ell$ !

On scales of the order of $\ell$, and below, one expects  deviations 
from the usual Newton's law.
Since  gravity experiments\cite{Hoyle_etal04}
 have  confirmed the standard Newton's 
law 
down to scales of the order $0.1$ mm, this implies  
\beq
\ell\lesssim 0.1 {\rm mm},
\eeq
and thus $M_{(5)}\gsim 10^8$ GeV.

Although 
the above results  apply to  linearized gravity, 
other works, based on second order calculations or numerical gravity,
have confirmed   the recovery 
of standard gravity on scales larger than $\ell$.
However, the behaviour of black holes in  the Randall-Sundrum model 
 might significantly 
deviate from the standard picture. Indeed, 
inspired by the AdS/CFT correspondence, it has been conjectured that 
Randall-Sundrum black holes  should {\it evaporate classically}, 
or, in other words,  be classically unstable.
The underlying argument is that the five-dimensional classical solutions 
should correspond to {\it quantum-corrected} 
four-dimensional black hole solutions, 
of a  conformal field theory (CFT) coupled to gravity\cite{tanaka,efk}. 
Since there are many  CFT degrees of freedom 
 into which the black hole can radiate, its 
 life time is shorter than that of  a standard black hole: 
\beq
\tau \simeq 10^2\left({M/ M_\odot}\right)^3\left(\ell/1{\rm mm}\right)^{-2}
{\rm years}.
\eeq

\section{Homogeneous brane cosmology}
Let us  now discuss the cosmology of a brane embedded in a five-dimensional 
bulk spacetime.

\subsection{The model}
As in standard cosmology, {\it homogeneity}
 and {\it isotropy} 
along  the three ordinary spatial dimensions are assumed. 
The bulk spacetime is thus required to satisfy  the {\it cosmological symmetry}, 
i.e.  one can foliate the bulk 
 with maximally symmetric three-dimensional surfaces. Note that this is 
in complete   analogy with {\it the spherical symmetry}, associated with 
 maximally  symmetric two-dimensional surfaces
in a 4D spacetime. 

In addition to the three ordinary spatial dimensions, spanning the 
homogeneous and isotropic surfaces, one  introduces a time 
coordinate $t$ and a spatial coordinate $y$ for the extra dimension. The 
cosmological symmetry implies that the metric components depend only 
 on $t$ and $y$.
It is convenient to work in  a Gaussian Normal (GN)
 coordinate system, in which the brane is always located at $y=0$ and 
the five-dimensional metric takes   the form 
\begin{equation}
\label{GN}
ds^2=- n^2(t,y)dt^2+a^2(t,y)d\Sigma_k^2+dy^2,
\end{equation}
where $d\Sigma_k^2$ is the metric for the maximally symmetric 
three-surface ($k=0,\pm 1$). 
Note that, in closer analogy with the spherical symmetry mentioned above, 
another possibility would be to choose a coordinate system in which  
the metric reads 
\beq
\label{spherical}
ds^2=- n^2(t,r)dt^2+b^2(t,r) dr^2+ r^2 d\Sigma_k^2.
\eeq

To obtain the equations governing the 
cosmological evolution, one substitutes the ansatz (\ref{GN}) into
 the five-dimensional Einstein equations
\begin{equation}
G_{AB}+\Lambda g_{AB}=\kappa^2 T_{AB}
\end{equation}
where  the energy-momentum tensor, assuming the  bulk empty, 
is only due to the brane matter and  thus given by
\begin{equation}
T_A^B=Diag(-\rho_b(t), P_b(t), P_b(t), P_b(t), 0)\, \delta(y),
\end{equation}
where $\rho_b$ and $P_b$ are respectively the total energy density and 
pressure in the brane. 
The five-dimensional Einstein 
equations can be solved explicitly~\cite{bdel99} and 
one  gets a solution for the metric components 
$n(t,y)$ and $a(t,y)$, in terms of $\rho_b(t)$ and $P_b(t)$, defined 
up to an integration constant. 

\subsection{The cosmological evolution in the brane}
On the brane, the metric is given by 
\beq
ds_b^2=-n_b(t)^2 dt^2 +a_b(t)^2 d\Sigma_k^2,
\qquad n_b(t)\equiv n(t,0), \quad a_b(t)\equiv a(t,0).
\eeq
It can be shown that the scale factor $a_b(t)$ satisfies  
the {\it modified 
Friedmann equation}~\cite{bdel99}:
\begin{equation}
\label{fried}
H_b^2\equiv {\dot a_b^2\over a_b^2}={\Lambda\over 6}+{\kappa^4\over 36}
\rho_b^2+{\C\over a_b^4}-{k\over a_b^2},
\end{equation} 
where $\C$ is an integration constant.
It can also be shown that, for an empty bulk, the  
usual  conservation equation  holds, which implies
\begin{equation}
\dot\rho_b+3H_b(\rho_b+P_b)=0.
\end{equation}

 For $\Lambda=0$ and $\C=0$,  the bulk is 5-D Minkowski and  the cosmology 
is highly unconventional since the Hubble parameter is proportional to 
the brane energy density~\cite{bdl99}. This is incompatible 
with  the standard nucleosynthesis scenario, which depends sensitively 
 on 
the expansion rate. 

To obtain a viable brane cosmology scenario, the simplest way is to 
generalize the  Randall-Sundrum model
to cosmology\cite{cosmors}. In the 
static version of the previous section, the energy density of the 
``Minkowski'' brane 
was  $\rho_b=\sigma_{RS}\equiv 6\M^3/\ell $. This can be generalized to 
an FLRW brane by adding to the intrinsic tension $\sigma_{RS}$ 
the usual cosmological energy density
$\rho(t)$ so that the total energy density is given by 
\beq
\label{rho_b}
\rho_b(t)=\sigma_{RS}+\rho(t).
\eeq
Moreover, the bulk is assumed to be endowed 
 with a negative cosmological constant $\Lambda<0$, satisfying the 
constraint (\ref{rs_constraint}).

Substituting the decomposition (\ref{rho_b}) into
 the Friedmann equation  (\ref{fried}), one finds 
\begin{equation}
\label{bdl}
H_b^2={8\pi G\over 3} \rho +{\kappa^4\over 36}\rho^2
+{\C\over a_b^4}-{k\over a_b^2}.
\end{equation}
In the expansion in $\rho$, the constant term vanishes because of the 
constraint (\ref{rs_constraint}), whereas the coefficient of the linear 
term is the standard one because 
$
8\pi G\equiv \kappa^4\sigma/6$, as implied by 
 (\ref{rs_constraint}) and (\ref{planck}).
However, the Friedmann equation (\ref{bdl}) 
 is characterized by two new features:
\begin{itemize}
\item a $\rho^2$ term, which dominates at high
energy;
\item a radiation-like term,  ${\C/ a_b^4}$, usually called 
{\it dark radiation}.
\end{itemize}
The cosmological evolution undergoes  a transition from a high 
energy regime, $\rho\gg \sigma$, characterized by 
an unconventional behaviour of the scale factor, into a 
low energy regime which reproduces our standard cosmology. For 
$\C=0$, $k=0$ and an equation of state $w=P/\rho=const$, one can solve 
analytically the evolution equations and one finds
\begin{equation}
a(t)\propto t^{1/q}\left(1+{q\, t\over 2\ell}\right)^{1/q},
\qquad q=3(1+w).
\end{equation}
One clearly sees the transition, at the epoch $t\sim \ell$, between 
the early, unconventional, evolution $a\sim t^{1/q}$ and the standard
evolution $a\sim t^{2/q}$.

In order to 
be compatible with the nucleosynthesis scenario,  
the high energy regime, where the cosmological evolution is unconventional, 
must take place before 
nucleosynthesis. This requires $\sigma^{1/4} \gsim 1$ MeV, and since 
$\sigma= 6/(\kappa^2\ell)=6 M_5^6/M_P^2$, this gives the constraint
$M_5\gsim 10^4$ GeV. This 
is much less stringent than the constraint 
from   small-scale gravity experiments, 
which presently require $\ell \lsim 0.1$ mm and $M_5 \gsim 10^8$ GeV.
As will be detailed in the next section, another observational constraint
applies to the dark radiation constant $\C$.

\subsection{Another point of view}
If,  instead of the GN ansatz (\ref{GN}) for the metric, 
one starts from the metric (\ref{spherical}), in analogy with 
the spherical symmetry, one can use  the generalization of the Birkhoff 
theorem, which states that a vacuum spherical symmetric solution of 
Einstein's equation is necessarily static and its geometry is  Schwarschild: 
the 5D vacuum cosmologically symmetric solution of 5D Einstein's 
equations with a (negative) cosmological constant is necessarily 
static and corresponds to the 
 AdS-Schwarzschild metric in five dimensions:
\begin{equation}
\label{ads}
ds^2=-f(R)dT^2+{dR^2\over f(R)}+R^2 d\Sigma_k^2,
\quad
f(R)=k+{R^2\over \ell^2}-{\C\over R^2}, \quad k=0,\pm 1.
\end{equation}
In this coordinate system, the brane is {\it moving} and 
 the modified Friedmann equation obtained above can be recovered from  
the junction conditions (\ref{junction})~\cite{kraus,ida}.

\section{Dark radiation}

So far, the bulk has been assumed to be {\it strictly empty}, apart from 
the presence of the brane.
However, the fluctuations of brane matter generate bulk 
gravitational waves. Equivalently, the scattering of brane 
particles  produce bulk gravitons 
($
\psi+{\bar \psi}\rightarrow G
$). 
Therefore, a realistic model must take into account the presence
of these bulk gravitons
that are emitted by the brane and then propagate in 
the bulk.

The rate of emission of these gravitons by the brane can be computed 
explicitly 
when the brane matter is in thermal equilibrium (with a temperature 
$T$). The corresponding energy loss rate is given by\cite{lsr02,hm01} 
\begin{equation}
\label{emission}
\dot\rho+4H\rho=-{315\, \over 512\, \pi^3}
\, \hat g \, \kappa^2\,  T^8,
\end{equation}
with an effective number of degrees of freedom given by  the following 
weighted sum of  scalar, vector and fermionic degrees of 
freedom:
\beq
\hat g=(2/3)g_s+4g_v +g_f.
\eeq

The energy transfer from the brane into the bulk modifies the cosmological evolution 
of the brane, on the one hand because the evolution of the energy density of the brane 
is modified, on the other hand because the bulk geometry is affected by the gravitons.
One can treat self-consistently such an energy transfer by using a five-dimensional 
generalization of the Vaidya solution.

\subsection{Vaidya model}
Let us start with the following metric, which generalizes the Vaidya metric to a five-dimensional 
bulk with a (negative) cosmological constant:
\beq
\label{vaidya}
ds^2=-\left(k+{r^2\over \ell^2}-{{\cal C}(v)\over r^2}
\right)dv^2+2drdv+r^2 d{\bf x}^2,
\eeq
where $v$ is a null coordinate and ${\cal C}(v)$ is a function that generalizes the constant 
${\cal C}$ of the AdS-Schwarzschild metric. If ${\cal C}$ is  constant, one recovers
(\ref{ads}) via a change of coordinate.

In the general case, the above metric (\ref{vaidya}) is a solution of the five-dimensional 
Einstein equations, with a null bulk energy-momentum tensor, i.e. of the form
\beq
T_{ab}=\psi\,  k_a k_b, \quad k_ck^c=0.
\eeq
One can then show that the cosmological brane evolution is completely determined by 
the following coupled system\cite{lsr02}
\def\hH{{\hat H}}
\def\hrho{{\hat \rho}}
\def\t{{\hat t}}
\def\hC{{\hat{\cal C}}}
\def\F{{\sigma}}
\begin{eqnarray}
&&{d\hrho\over d\t}+4\hH\hrho=-\alpha\hrho^2,\cr
&&\hH^2=2\hrho+\hrho^2+{\hC\over a^4},\cr
&&{d\hC\over d\t}=2\alpha a^4\hrho^2\left(1+\hrho-\hH\right),
\nonumber
\end{eqnarray}
for  the dimensionless quantities $\hrho=\rho/\sigma_{RS}$, $\t=\ell t$, $\hH=H\ell$ and $\hC=\C \ell^2$.
The first two equations, the energy non-conservation and the Friedmann equation, are a consequence of the junction conditions whereas the third equation follows from the Einstein equations.
This system can be solved analytically \cite{leeper} and one finds that, in the low energy regime, 
${\cal C}$ tends toward a constant, which means that the production of bulk gravitons can be 
neglected. 
Although this model is rather nice (and has been generalized to non-$Z_2$ symmetric branes \cite{asym_vaidya}), it is not realistic because it implicitly
assumes that the bulk  gravitons must be emitted radially, which is not the case.

\subsection{More realistic treatment}
After their emission, the gravitons propagate freely in the bulk 
where they follow geodesic trajectories. As illustrated in Fig.~3, some 
of these 
gravitons (in fact many)  tend to come back onto 
the brane and  bounce off it.
\begin{figure}[t]
  \begin{center}
    \includegraphics[height=13pc]{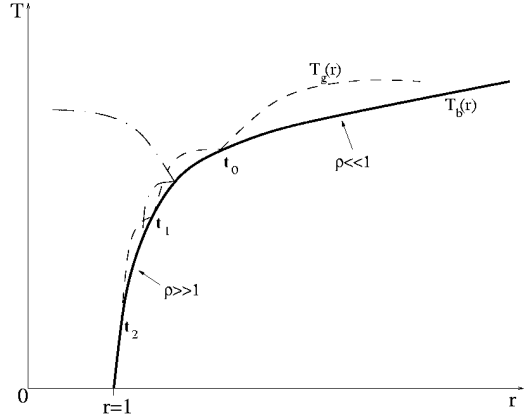}%%%%% ``13pc'' is just the example.
  \end{center}
  \caption{Examples of graviton trajectories in the bulk (dashed line). 
The brane 
trajectory, with relativistic matter, is also shown (continuous line).}
\end{figure}
All these gravitons contribute to an effective bulk energy-momentum
tensor, which can be written as 
\begin{equation} 
\T_{AB}=\int d^5p \ \delta\left(p_Mp^M\right)\sqrt{-g}\,f\, p_Ap_B, 
\label{T_integ},
\end{equation}  
where $f$ is the phase space distribution function.

 From the 5D Einstein equations,
one can derive effective 4D Einstein equations~\cite{sms99}, 
which in the homogeneous 
case yield 
\begin{itemize}
\item the  Friedmann equation
\begin{equation}
H^2 =   
{8\pi G \over 3}\left[  
\left(1+{\rho\over 2\sigma}\right)\rho+ \rho_{\rm{D}}  \right],  
\label{Hubble} 
\end{equation} 
\item the non-conservation equation for brane matter, 
which must be identified with (\ref{emission}),
\begin{equation} 
\dot{\rho}+3\,H\,\left(\rho+p\right)=2\,{\T}_{RS}\,n^R\,u^S\, ,
\end{equation} 
where $n^A$ is the unit vector normal to the  brane and 
$u^A$ its velocity in the bulk;
\item the non-conservation equation for the ``dark'' component
$\rho_{\rm{D}}$ (which includes all effective contributions from the bulk):
\begin{equation}
\label{dark}
 \dot \rho_{\rm{D}}+4H\rho_{\rm{D}} 
  =-2\left(1+{\rho\over\sigma}\right)\T_{AB} u^A n^B   
  -2 H\ell \, \T_{AB} n^A n^B\,. 
\end{equation} 
\end{itemize}
On the right hand side of this last equation, we find two terms 
involving the bulk energy-momentum tensor: 
the first term, due to the energy flux from the brane 
into the bulk, contributes positively and thus increases the amount of 
dark radiation whereas the second term, due to the pressure along 
the fifth dimension, decreases the amount of dark radiation.
These terms can be estimated numerically~\cite{ls03}. A striking 
property is  that the 
 gravitons  coming  back onto 
the brane and  bouncing off it  give a significant contribution 
to the 
transverse pressure effect, which almost, although not quite, compensates
the flux effect. 
The evolution of the dark radiation, or rather its ratio with respect 
to the brane radiation density
$\epsilon_D\equiv \rho_D/\rho$,  is plotted on Fig.~4.
\begin{figure}[t]
  \begin{center}
    \includegraphics[height=20pc,angle=-90]{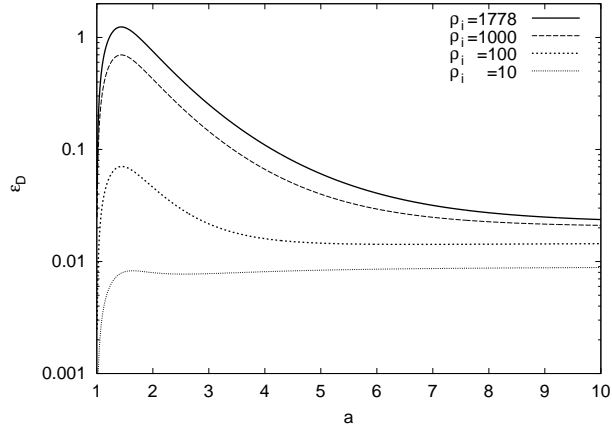}%%%%% ``13pc'' is just the example.
  \end{center}
  \caption{Evolution of the  ratio  $\epsilon_D=\rho_D/\rho$ for different 
values of the initial energy density on the brane 
$\rho_i$ (in units of $\sigma^4$)~\cite{ls03}.}
\end{figure}
At late times, i.e. far in the low energy regime, the 
ratio reaches a plateau, because the right hand side of (\ref{dark}) becomes
negligible. The dark component then scales exactly like radiation.

\subsection{Observational constraints}
The computed amount of dark radiation can be confronted to observations. 
Indeed, since 
dark radiation behaves as radiation, it must satisfy the  
 nucleosynthesis  constraint on the number of {\it additional 
relativistic degrees of freedom}, usually expressed in terms 
of the extra number of light neutrinos $\Delta N_\nu$.
 The relation between $\Delta N_\nu$ and $\epsilon_D$ is given by
\begin{equation}
\epsilon_D={7\over 43}\left({g_*\over g_*^{\rm nucl}}\right)^{1/3}
\Delta N_\nu,
\end{equation}
where $g_*^{\rm nucl}=10.75$ is the number of degrees of freedom at
nucleosynthesis (in fact before the electron-positron annihilation).
Assuming $g_*=106.75$ (standard model), this gives
$\epsilon_D\simeq 0.35 \Delta N_\nu$. 
The typical constraint from nucleosynthesis 
$\Delta N_\nu \lesssim 0.2$
thus implies
\begin{equation} 
\epsilon_D\equiv {\rho_D\over\rho_r} \lesssim 0.03 \left({g_*\over 
g_{*}^{\rm nucl}}
\right)^{1/3},
\end{equation}
which gives $\epsilon_D \lesssim 0.09$ with the degrees of freedom 
 of the standard model.

\section{Anisotropic brane cosmology}
The homogeneous and isotropic brane cosmology is in fact very 
simple because  of the generalized Birkhoff's theorem mentioned 
earlier. But, when the cosmological symmetry are relaxed, things 
become rather  difficult because the bulk geometry is no longer 
Schwarzschild-AdS. 
As a first step towards the general case, it is instructive to study 
 configurations where the  
 cosmology in the brane is homogeneous 
but anisotropic, e.g. of the Bianchi 
I type with a metric of the form
\beq
ds^2_b=-d\tau^2+\sum_{i=1}^3a_i^2(\tau)(dx^i)^2.
\eeq
Although many works in the literature 
have been devoted to this subject, most of them use  the effective
four-dimensional equations projected on the brane. It is a  more 
challenging task to solve the 5D Einstein equations for the bulk as well, 
starting e.g. from an ansatz of the form 
\beq
 ds_{\rm bulk}^2=-{n^2(t,y)}dt^2+\sum_{i=1}^3
{a_i^2(t,y)}({dx^i})^2+dy^2.
\eeq
Assuming that the metric is separable, it turns out that  
explicit  solutions have been found. They are given by~\cite{flsz04}
\begin{eqnarray}
ds^2=\sinh^{1/2}(4y/\ell)\left[\right. &-& \tanh\left(2y/\ell\right)^{2 q_0} dt^2
\cr
&+&\left.\sum_i \tanh\left(2y/\ell\right)^{2 q_i}t^{2p_i}\left(dx^i\right)^2\right]
+dy^2,
\end{eqnarray}
where the seven coefficients $q_\mu$ and $p_i$ must satisfy the constraints
\beq
\sum_{\mu=0}^3 q_\mu=0, \quad \sum_\mu q_\mu^2={3\over 4}, \quad 
\sum_{i=1}^3 p_i=1 , 
\quad \sum_i p_i^2=1, \quad \sum_i q_i\left(p_i+1\right)=0.
\eeq

In general, a brane embedded in an anisotropic bulk spacetime must
contain matter with {\it anisotropic stress}, 
because of the junction conditions, which can here be separated into two parts: 
\bi
\item  isotropic part:
\beq
n^{-1}\dot y_b  \left. \dot{A} \right|_b + 
\sqrt{1+\dot y_b^2 } \, \left. {A'} \right|_b = 
\frac{\kappa^2}{6} \rho_b \label{Cond1},
\eeq
\item anisotropic part:
\beq
n^{-1}\dot y_b \left. \dot{B}_i \right|_b +
\sqrt{1+\dot y_b^2}\; \left. {B'}_i \right|_b =
\frac{\kappa^2}{2}\pi_i, \label{Cond2}
\eeq
\ei
where $\pi_i$ is the anisotropic pressure in the brane, and using the
notation $3A\equiv\ln(a_1a_2a_3)$ and 
$B_i\equiv\ln a_i -A$.
Note that the brane position $y_b$ is not assumed to be fixed 
here: in this sense
the coordinate system is not Gaussian Normal.
Interestingly, the above  solutions include 
a particular bulk geometry, for $q_0=\pm \sqrt{3}/4$, in which 
one can embed a moving 
brane with  perfect fluid as  matter, i.e. $\pi_i=0$. For this particular case, 
the effective cosmological equation of state
$P_{\rm eff}/ \rho_{\rm eff}$ is negative but goes to zero 
at late times.

\section{Brane-worlds with generalized gravity theories}
So far, we have used five-dimensional Einstein gravity for the bulk. However, one 
might envisage more general gravity theories to describe the bulk space-time. 
One possibility is to take into account higher order curvature terms in the five-dimensional
action. It turns out that there is a particular combination of the second order curvature 
terms, called the Gauss-Bonnet term, which yields well-behaved  equations of motion. 
The five-dimensional action with a Gauss-Bonnet term reads\footnote{
It is interesting to note that this Einstein-Gauss-Bonnet theory has been introduced 
in a six-dimensional bulk in order to recover Einstein gravity on a codimension 2 brane \cite{Bostock:2003cv}. 
This result has been extended to any even dimension in the more 
general context of Lovelock gravity theories \cite{charmousis_zegers}.}
\beq
{\cal S} = \frac{1}{2\kappa_5^2} \int d^5x \sqrt{-\, g_5}
\left[-2\Lambda_5+ {\cal R} 
%\right. \nonumber\\  &&\left.~{}
+\alpha\, \left({\cal R}^2-4 {\cal R}_{ab}{\cal
R}^{ab}+ {\cal R}_{abcd}{\cal R}^{abcd}\right) \right] 
\nonumber\\
%&&~{} - \int_{\rm brane} d^4x\, \sqrt{-g}\, \sigma\,,\nonumber
\eeq

All the steps discussed previously can be revisited in this more general context. 
One can find a Minkowski brane if the brane tension is adjusted to the value
\beq
\kappa_5^2\, \sigma = 2(3-\beta)/\ell
\eeq
where $\beta\equiv 4\alpha/\ell^2$.
The effective four-dimensional gravitational constant is given by 
\beq
\kappa_4^2= \frac{\kappa_5^2}{\ell \, (1+\beta)}\,.
\eeq

In the cosmological context, the modified Friedmann equation is 
now given by \cite{charmousis_dufaux}
\beq
\kappa_5^2(\rho+\sigma) =
{2\over\ell}\sqrt{1+{H^2\ell^2}}\, \left[3+\beta\left(
{2H^2\ell^2-1}\right)\right].
\eeq
Assuming that the Gauss-Bonnet term represents a small correction to the Einstein-Hilbert term, 
i.e. $\beta\ll 1$, this Friedmann equation exhibits  three different regimes: 
\bi
\item a Gauss-Bonnet regime for $H\ell \gg \beta^{-1}$, during which 
\beq
H^2\approx \left[
{\kappa_{_5}^2 \over 4\beta\ell^2}\, \rho \right]^{\!2/3},
\eeq 
\item a five-dimensional Einstein regime for $1\ll H\ell\ll \beta^{-1}$, with the behaviour
\beq
H^2\approx
{\kappa_{_5}^2
\over 36}\, \rho^{2},
\eeq
\item finally, the ordinary four-dimensional Einstein regime for $H\ell \gg 1$, characterized 
by the usual Friedmann law 
\beq
H^2\approx {\kappa_4^2 \over 3}\, 
\rho.
\eeq

\ei

\section{Brane inflation}
In brane cosmology,  the famous horizon problem  is much less 
severe than in standard cosmology, 
because the gravitational horizon, associated with the signal propagation
in the bulk, can be much 
bigger than the standard photon horizon, associated with the 
signal propagation on the brane\cite{cl01}.
However, it  is still alive
because the 
energy density on the brane is limited by the  Planck limit $\rho\sim \M^4$.
Thus, one must still invoke inflation, altough alternative ideas 
based on the collision of branes\cite{kost}
 have been actively  explored\footnote{Note, however, that the 
generation of a quasi-scale-invariant fluctuation spectrum, as required 
by observations, remains problematic because the fifth dimension goes to zero  when two $Z_2$-symmetric
branes collide and the evolution of perturbations is then ill-defined. By contrast, the collision of branes 
that are not both $Z_2$-symmetric is well-behaved and precise conservation laws can be derived~\cite{Langlois:2001uq}.}.

The simplest way 
to get inflation in the brane\footnote{We do not consider here models, also called brane inflation, where 
the inflaton is the distance between two branes in relative motion  with respect to each other. These models are based  on 
an effective four-dimensional approach.  In the present context, where the self-gravity of the brane is 
essential, a four-dimensional approach does not apply~\cite{Langlois:2002hz}
 except in the low-energy limit~\cite{ks}.}
is to detune the brane tension from 
its Randall-Sundrum value (\ref{rs}) in order to obtain  a 
net effective four-dimensional cosmological constant that is positive. This
 leads to  exponential expansion on the brane.
In the GN coordinate system, the metric 
is separable and  
can be written as 
\beq
ds^2= \A(y)^2 \left(-dt^2+e^{2H t} d{\vec x}^2\right)+dy^2,
\label{dS}
\eeq
with 
\beq
\A(y)=   \cosh\mu y-\left(1+{\rho\over\sigma}\right) \sinh\mu|y|.
\label{A}
\eeq

As in the Randall-Sundrum case, the linearized Einstein equations 
for the tensor modes lead to a {\it separable} wave equation.
The shape along the fifth dimension 
of the corresponding massive modes is governed by 
the Schr\"odinger-like equation 
\beq
\label{SE}
 {d^2\Psi_m\over dz^2} - V(z)\Psi_m =-m^2 \Psi_m \,,
\end{equation}
after introducing   the new variable  
$z-z_b=\int_0^y d\tilde y/\A(\tilde y)$ 
(with $z_b=H^{-1}\sinh^{-1}(H\ell)$)
and the new function
 $\Psi_m= \A^{-1/2}u_m(y)$.
The potential is given by 
\beq
V(z)= {15H^2 \over 4\sinh^2(H z)} +
{{9\over4}}H^2
- {3\over\ell}\left(1+{\rho\over\sigma}\right) \delta(z-z_{\rm b}) \,
\eeq
and plotted in Fig.~5. 
\begin{figure}[t]
  \begin{center}
    \includegraphics[height=12pc,angle=0]{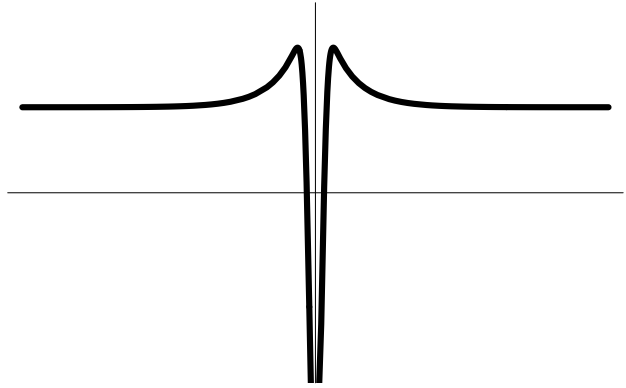}%%%%% 
  \end{center}
\caption{Potential for the graviton modes in a de Sitter brane}
\end{figure}
In constrast with the Randall-Sundrum potential, the potential 
goes asympotically to the non-zero value $9H^2/4$. This indicates 
 the presence 
of a gap
between the zero mode ($m=0$) and the continuum of Kaluza-Klein modes 
($m>3H/2$).

In practice, inflation is not strictly de Sitter but the de Sitter case 
discussed above is a good approximation when $\dot H\ll H^2$. To get 
``realistic'' inflation in the brane, two main approaches 
 have been considered: 
either to assume  a five-dimensional  scalar field which induces inflation
in the brane\cite{sasaki}, or to suppose  
a four-dimensional scalar field confined on the
 brane\cite{mwbh99}.

In the latter case, the cosmological evolution during inflation 
is obtained by substituting the energy density 
$\rho_\phi=\dot\phi^2/2 + V(\phi)$
in the modified Friedmann equation (\ref{bdl}). 
For slow-roll inflation, this can be approximated by 
\beq
H^2\simeq {8\pi G\over 3}\left(1+{V\over 2\sigma}\right)V.
\eeq
Interestingly, because of the modified Friedmann equation, 
new features appear at  high energy ($V> \sigma$):
the slow-roll 
conditions are changed and,  because the Hubble parameter is bigger 
than the standard value, yielding a higher friction on the scalar field,
inflation can occur  with potentials usually too steep to 
sustain it~\cite{cll00}.

The scalar and tensor spectra generated during inflation driven by 
a brane scalar field have also been computed~\cite{mwbh99,lmw00} (although 
there might be some subtleties for the scalar modes \cite{Koyama:2004ap,Koyama:2005ek}). 
They are modified 
with respect to the standard results, according to the expressions
\beq
P_S=P_S^{\rm(4D)}\left(1+{V\over 2\sigma}\right)^3,
\qquad
P_T=P_T^{\rm (4D)}F^2(H\ell),
\eeq
with 
\beq
F\!\left(x\right) =\left\{ \sqrt{1+x^2} - x^2 \ln \left[ {1\over
x}+\sqrt{1+{1\over x^2}} \right] \right\}^{\!\!-1/2}
\!.
\eeq
At low energies, i.e. for $H\ell\ll 1$,  $F\simeq 1$ and 
one  recovers exactly the usual four-dimensional result 
but at higher energies the multiplicative factor $F$ provides an 
{\it enhancement} of the gravitational wave spectrum amplitude 
with respect to the four-dimensional result: 
  $F\simeq {(3/2)H\ell}\sim V/\sigma$ at very high energies, 
i.e. for $H\ell\gg 1$.
Nevertheless, comparing  
this with the amplitude of the scalar spectrum,
one finds that, at high energies ($\rho\gg\sigma$), the {\it 
tensor over scalar 
ratio is in fact suppressed} with respect to the four-dimensional ratio.

These results have been extended to the case of the Einstein-Gauss-Bonnet theory discussed in 
the previous section\cite{Dufaux:2004qs}. The scalar spectrum is given by
\beq
{\cal P}_{\rm S} 
={\cal P}_{\rm S}^{_{\rm 4D}}\,  G_\beta^2(H\ell)
\eeq
with
\beq
G^2_\beta(x)=\left({3(1+\beta) x^2 \over 2\sqrt{1+x^2}(3-\beta
+2\beta x^2)+2(\beta-3)}\right)^{3}
\eeq
whereas the tensor spectrum is given by 
\beq
{\cal P}_{\rm T} 
={\cal P}_{\rm T}^{_{\rm 4D}}\,  F_\beta^2(H\ell)
\eeq
with
\beq
F_\beta^{-2}(x)=\sqrt{1+x^2}-\left(\frac{1-\beta}{1+\beta}
\right)x^2 \sinh^{-1}\frac{1}{x}\,. \label{F}
\eeq
All these results give the amplitude of the pertubations during inflation. 
A difficult question  is to determine how the perturbations  will evolve during the 
subsequent cosmological phases, the radiation and matter eras.

\section{Cosmological perturbations}
A crucial test  for brane cosmology is the  confrontation  with cosmological 
observations, in particular the  CMB fluctuations.  Although the {\it 
primordial}
power spectra for scalar and tensor perturbations have been computed, the 
subsequent evolution  of the cosmological perturbations is non trivial 
and has not been fully solved yet. Indeed, 
in contrast with standard cosmology where the 
evolution of cosmological perturbations can be reduced to  ordinary 
differential equations for the Fourier modes, the evolution equations in brane
cosmology are partial differential equations with two variables: the time 
and the fifth coordinate. 
Another delicate point is to specify the boundary conditions, both in time 
and space. 

An instructive, although limited, approach for  brane cosmological 
perturbations is the brane point of view, based on the 4D effective Einstein
equations on the brane, usually written in the form~\cite{sms99}
\beq
\label{einstein_4d}
G_{\mu\nu}+\Lambda_4\,  g_{\mu\nu}=8\pi G \tau_{\mu\nu}+\kappa^2 \Pi_{\mu\nu}
-E_{\mu\nu},
\eeq
where $\tau_{\mu\nu}$ is the brane energy-momentum tensor, 
$\Pi_{\mu\nu}$ is a tensor depending quadratically on  $\tau_{\mu\nu}$ (which  
gives the $\rho^2$ term of the Friedmann equation in the homogeneous case), and 
$E_{\mu\nu}$, which corresponds to the dark radiation in the 
homogeneous case, is the projection on the brane of the bulk Weyl tensor.

It is then a straightforward exercise, starting from (\ref{einstein_4d}),
 to write  down explicitly the perturbed effective
Einstein equations on the brane, which will look exactly as the 
four-dimensional ones for the geometrical part but with extra terms 
due to $\Pi_{\mu\nu}$ and $T^{Weyl}_{\mu\nu}$. One thus gets
 equations relating  the perturbations of the metric 
to  the matter  perturbations {\it and} the perturbations of 
the projected Weyl tensor, which formally can be assimilated to a virtual 
fluid, with  corresponding (perturbed) energy density $\rho_E+\delta\rho_E$,
pressure $P_E+\delta
P_E={1\over3}(\rho_E+\delta\rho_E)$ and anisotropic pressure
\cite{Langlois:2000ph}.
The contracted Bianchi identities ($\nabla_\mu G^\mu_\nu=0$)
and energy-momentum conservation for matter on the brane 
($\nabla_\mu \tau^\mu_\nu=0$) ensure, using Eq.~(\ref{einstein_4d}), that
\begin{equation}
\nabla_\mu E^\mu_\nu = \kappa^4\,\nabla_\mu\Pi^\mu_\nu \,.
\end{equation}
In the background, this tells us that $\rho_\E$ behaves like radiation, 
as we knew already,
and for the first-order perturbations, one finds
 that the effective energy of the projected Weyl
tensor is conserved independently of the quadratic energy-momentum
tensor. The only interaction is a momentum transfer.
 
It is also possible to construct~\cite{lmsw00} gauge-invariant variables 
corresponding to the curvature perturbation on hypersurfaces of
uniform density, both for the brane matter energy density 
and for the total effective energy density (including the quadratic terms 
and the Weyl component). These quantities are extremely useful because 
their evolution on scales larger than the Hubble radius can be solved easily.
However, their connection to the large-angle 
CMB anisotropies involves the knowledge of anisotropic stresses due to the 
bulk metric perturbations. 
This means that  {\it for a quantitative prediction 
of  the CMB anisotropies, even at large scales,  one needs to determine 
the evolution of the bulk perturbations}.

In summary, one can  obtain a set of equations for the brane linear 
perturbations,  
where one recognizes the ordinary cosmological  equations 
but modified by   two  types of corrections:
\begin{itemize}
\item modification of the  homogeneous background coefficients due to the 
additional $\rho^2$ terms in the Friedmann equation.   These  corrections are  
negligible in the low energy  regime $\rho\ll\sigma$;
\item presence of source terms in the  equations. 
These terms come from the bulk perturbations and cannot be determined solely 
from the evolution inside the brane. To determine them, one must solve 
the full problem in the bulk (which also means to specify some initial 
conditions in the bulk). 
\end{itemize}
Most of the recent works~\cite{tensor} studying the post-inflation evolution 
of brane cosmological perturbations
have concentrated on   tensor modes, which are simpler because they are not 
coupled, like scalar modes, to brane matter fluctuations.

 \section{Conclusions}
In this contribution, I have presented some aspects of brane-world 
models, covering both the (static) Randall-Sundrum model and its cosmological
extensions. Due to lack of time/space, I have not discussed many 
other  interesting topics in the field. Examples are the brane cosmology 
of  models involving 
Gauss-Bonnet corrections; the induced gravity  models, where 
one includes a 4D Einstein-Hilbert action for the brane and which 
can lead to late-time cosmological effects mimicking dark energy.

There are still many open questions in brane cosmology. Even in the simplest 
set-up, discussed here, based on a cosmological extension of the 
Randall-Sundrum model, the evolution of cosmological perturbations has not
yet been solved, although some significant progress has been made.
 The situation is still more complicated  in more sophisticated models, 
involving a bulk scalar field and/or collision of branes. 
It must be emphasized that the predictions for the  cosmological 
perturbations, as observed in the CMB experiments, and their adequation 
with the present data, 
is a crucial test
for brane-world models for which the early universe is modified.
More direct tests of 
brane-world models involve 
 gravity experiments or  collider experiments. However, 
if the fundamental Planck mass is too high, such direct experiments
cannot see extra-dimensional effects and one must turn to cosmology to 
try to see indirect signatures from the early universe.

Another direction of  research is to make contact between the 
brane-worlds, which are still only  phenomenological models, 
and a fundamental theory like string theory.

\section*{Acknowledgments}
I would like to thank the organizers of the YKIS 2005 Conference, and especially Misao Sasaki, 
for inviting me to this very stimulating meeting and for their warm hospitality during my stay in
Kyoto. 

%%%%%%%%%%%%%%%%%%%%%%%%%%%%%%%%%%%%%%%%%%%%%%%%%%%%%%%%%%%%%
% Doing references:                                         %
%%%%%%%%%%%%%%%%%%%%%%%%%%%%%%%%%%%%%%%%%%%%%%%%%%%%%%%%%%%%%

%%%%%%%%%%
\end{document}